\documentclass[aps,prl,twocolumn,superscriptaddress]{revtex4}
\usepackage{epsf,graphicx}
\usepackage{amssymb}
\usepackage{amsmath}

\usepackage{color}
\usepackage{ulem}
\begin{document}

\title{Mutual information, quantum phase transition, and phase coherence in Kondo systems}
\author{Jian-Jun Dong}
\affiliation{Beijing National Laboratory for Condensed Matter Physics and Institute of
Physics, Chinese Academy of Sciences, Beijing 100190, China}
\affiliation{University of Chinese Academy of Sciences, Beijing 100049, China}
\author{Dongchen Huang}
\affiliation{Beijing National Laboratory for Condensed Matter Physics and Institute of
Physics, Chinese Academy of Sciences, Beijing 100190, China}
\affiliation{University of Chinese Academy of Sciences, Beijing 100049, China}
\author{Yi-feng Yang}
\email[]{yifeng@iphy.ac.cn}
\affiliation{Beijing National Laboratory for Condensed Matter Physics and Institute of
Physics, Chinese Academy of Sciences, Beijing 100190, China}
\affiliation{University of Chinese Academy of Sciences, Beijing 100049, China}
\affiliation{Songshan Lake Materials Laboratory, Dongguan, Guangdong 523808, China}
\date{\today}

\begin{abstract}
We propose a static auxiliary field approximation to study the hybridization physics of Kondo systems without the sign problem and use the mutual information to measure the intersite hybridization correlations. Our method takes full account of the spatial fluctuations of the hybridization fields at all orders and overcomes the artificial (first-order) phase transition predicted in the mean-field approximation. When applied to the two-impurity Kondo model, it reveals a logarithmically divergent amplitude mutual information near the so-called ``Varma-Jones" fixed point and a large phase mutual information manifesting the development of intersite phase coherence in the Kondo regime, with observable influences on physical properties. These highlight the importance of hybridization fluctuations and confirm the mutual information as a useful tool to explore the hybridization physics in Kondo systems.
\end{abstract}

\maketitle

Correlation between subsystems plays a key role in many-body quantum systems whose collective phenomena cannot be viewed as a simple addition of microscopic properties \cite{Anderson1972Science,Wen2019Springer}. The mutual
information, a key concept in the information theory, measures the statistical dependency between two random variables and may be used to probe the total amount of correlations between subsystems \cite{Thomas1991,Wolf2008PRL}. It has recently been introduced to identify phase transitions without explicit knowledge of the broken symmetry and the order parameter \cite{Matsuda1996IJTP,Wicks2007PRE,Wilms2011JSTAT,Wilms2012JSTAT,Iaconis2013PRB,Lau2013PRE,Casini2016JHEP,Hagymasi2015PRB,Toldin2019PRB,Hofmann2019PRB,Walsh2019PRL,Walsh2021PNAS,Valdez2017PRL,Bagrov2020SciRep}, but the concept has not been widely applied in strongly correlated electronic systems.

The Kondo systems are arguably one of the most well studied correlated systems. Theoretically, a pseudofermion representation is often used for the impurity spins and the underlying physics has been described by an effective hybridization between pseudofermions and conduction electrons \cite{Hewson1997,Coleman2015,Yang2017PNAS}. However, most studies focus on the mean-field approximation \cite{Doniach1977,Newns1987AdvPhys,Sire1993PRB,Zhang2000PRB,Coqblin2003PRB,Senthil2003PRL,Senthil2004PRB,Rech2006PRL,Bannon2016JHEP,Zhang2018PRB,Wugalter2020PRB,Coleman1987PRB,Jones1989PRB2,Jones1991PhysicaB} and ignore fluctuations of the hybridization fields that may give rise to important new physics as recently observed in ultrafast optical pump-probe experiments \cite{Liu2020PRL}. The mean-field theory has made false predictions such as an artificial (first-order) phase transition in multi-impurity Kondo systems \cite{Jones1991PhysicaB,Jones1989PRB2}. Efforts to take into account some thermal and quantum fluctuations have led to some effective low-energy theories but include only low order expansions of the hybridization due to analytical difficulties \cite{Auerbach1986PRL,Coleman2005PRB,Pepin2007PRL,Pepin2008PRB,Ohara2007PRB,Ohara2013JPSJ,Ohara2014JPSJ}. Numerical simulations \cite{Fye1989PRB,Fye1994PRL,Hallberg1995PRB,Silva1996PRL,Campo2004PRB,Bulla2008RMP,Zhu2011PRB,Bayat2012PRL,Bayat2014NC,Hu2019PRB,Raczkowski2019PRL,Cai2019arXiv,Hu2020arXiv} usually do not directly probe the hybridization fields due to the sign problem. A proper treatment of hybridization fluctuations is lacking, which severely limits our exploration of the richness of the hybridization physics.

In this work, we propose a static auxiliary field approximation to directly simulate the probabilistic distribution of the hybridization fields beyond the mean-field approach. We show that mutual information may be used as a tool to reveal some important aspects of the hybridization physics. The method allows us to capture full spatial correlations of the hybridization fields using the Monte Carlo sampling without the sign problem. As an example, when applied to the two-impurity Kondo model, it suppresses the artificial first-order phase transition, in good agreement with the exact numerical renormalization group (NRG) analysis \cite{Jones1987PRL,Jones1988PRL,Jones1989PRB1}. We find that the mutual information of the hybridization amplitude, calculated using the recently developed neural estimator, exhibits a logarithmic divergence with lowering temperature near the so-called ``Varma-Jones" fixed point, while that of the phase adopts a finite value in the Kondo regime but diminishes for large impurity distance, providing for the first time a measure of the intersite phase coherence. These have important influences on physical properties, in particular near the critical point. Our method can be easily extended to other models to provide useful insight on their hybridization physics.

We start with the two-impurity Kondo model,
\begin{equation}
H=\sum_{\mathbf{k}\sigma}\epsilon_{\mathbf{k}}c_{\mathbf{k}\sigma}^{\dag
}c_{\mathbf{k}\sigma}+J_\text{K}\sum_{i=1}^{2}\mathbf{s}_{i}\cdot
\mathbf{S}_{i}+J_\text{H}\mathbf{S}_{1}\cdot\mathbf{S}_{2}, \label{Hamiltonian}
\end{equation}
where $\epsilon_{\mathbf{k}}$ is the conduction electron dispersion, $J_{\text{K}}$ and $J_{\text{H}}$ are the Kondo and Heisenberg exchange interactions, respectively, $\mathbf{S}_{i}$ is the impurity spin located at $\mathbf{R}_{i}$, and $\mathbf{s}_{i}=\sum_{\alpha\beta}c_{i\alpha}^{\dag}\frac{\vec{\sigma}_{\alpha\beta}} {2}c_{i\beta}$ is that of conduction electrons. The model describes one of the simplest systems that feature strong electronic correlations and competition between different many-body ground states. In the magnetic limit ($ J_{\text{H}} \gg J_{\text{K}}$), the two impurities are locked into a spin singlet, while in the Kondo limit ($ J_{\text{K}} \gg J_{\text{H}}$), both are screened by conduction electrons. While the mean-field theory has predicted a first-order transition between two phases \cite{Jones1991PhysicaB,Jones1989PRB2}, NRG and the conformal field theory (CFT) suggest a crossover at finite temperatures and a ``Varma-Jones" fixed point at zero temperature for a conduction band with the particle-hole symmetry \cite{Jones1987PRL,Jones1988PRL,Jones1989PRB1,Affleck1992PRL,Affleck1995PRB}.

The hybridization physics is best seen in the Abrikosov pseudofermion representation, $\mathbf{S_{i}}=\sum_{\eta\gamma}f_{i\eta}^{\dag}\frac{\vec{\sigma}_{\eta\gamma}}{2}f_{i\gamma}$. The Kondo and Heisenberg terms can be decomposed using the standard Hubbard-Stratonovich factorization, resulting in a bilinear action of pseudofermions and conduction electrons that are only coupled through fluctuating background auxiliary fields,
\begin{equation}
S=\beta\sum_{i,n}(\frac{J_{\text{K}}\left\vert V_{i,n}\right\vert ^{2}} {2}+\frac{J_{\text{H}}\left\vert \chi_{n}\right\vert ^{2}}{4})-\beta\sum_{i=1}^{2}\lambda_{i}+S_{1},
\label{S1}
\end{equation}%
where $V_{i,n}$ and $\chi_{n}$ are the auxiliary fields in Matsubara frequency ($i\omega_{n}$) representing the Kondo hybridization and intersite magnetic correlation, respectively. $S_{1}=\sum_{nm\sigma}\Psi_{n\sigma}^{\dag }\left(  O_{nm}-\operatorname*{i}\omega_{n}\delta_{nm}\right)  \Psi_{m\sigma}$, with $\Psi_{n\sigma}=[c_{\mathbf{k}_{1}\sigma n},\cdots,c_{\mathbf{k}_{N_{0}}\sigma n},f_{1\sigma n},f_{2\sigma n}]^{\text{T}}$ and $N_{0}$ being the number of $\mathbf{k}$ points in the Brillouin zone of conduction electrons. The matrix $O_{nm}$ is
\begin{equation}
O_{nm}=\left[
\begin{array}
[c]{ccccc}%
\epsilon_{\mathbf{k}_{1}} & \cdots & 0 & V_{1,n-m}^{1} & V_{2,n-m}^{1}\\
\vdots & \ddots & \vdots & \vdots & \vdots\\
0 & \cdots & \epsilon_{\mathbf{k}_{N_{0}}} & V_{1,n-m}^{N_{0}} &
V_{2,n-m}^{N_{0}}\\
\overline{V}_{1,m-n}^{1} & \cdots & \overline{V}_{1,m-n}^{N_{0}} & \lambda_{1}
& \frac{J_{H}\chi_{n-m}}{2}\\
\overline{V}_{2,m-n}^{1} & \cdots & \overline{V}_{2,m-n}^{N_{0}} & \frac
{J_{H}\overline{\chi}_{m-n}}{2} & \lambda_{2}
\end{array}
\right],\nonumber
\end{equation}
where $V_{i,n-m}^{j}=\frac{J_{\text{K}}\operatorname{e}^{\operatorname*{i} \mathbf{R}_{i}\mathbf{\cdot k}_{j}}V_{i,n-m}}{2\sqrt{N_{0}}}$. $\lambda_i$ is the Lagrange multiplier for the constraints $n^i_f=1$ and takes a real value after a Wick rotation \cite{Zhou2017RMP}.

\begin{figure}[ptb]
\begin{center}
\includegraphics[width=8.6cm]{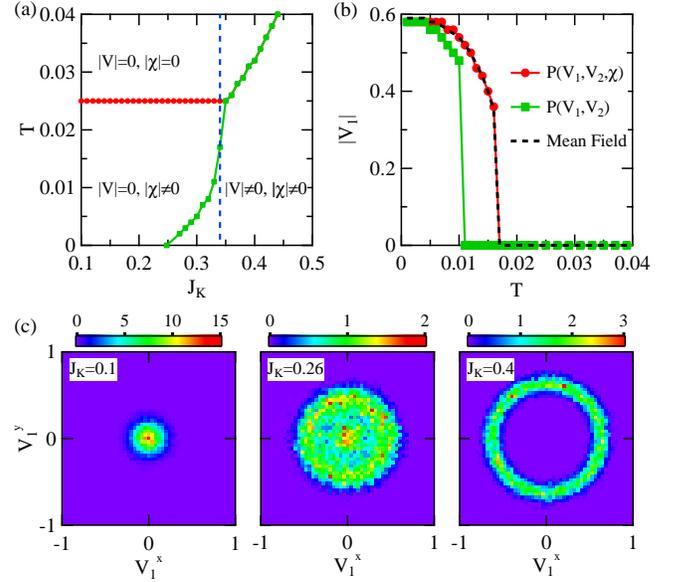}
\end{center}
\caption{(a) The mean-field phase diagram. (b) Comparison of the mean-field $|V_1|$ and that derived from the peak position of $p(  V_{1},V_{2}, \chi)$ and $p(  V_{1},V_{2})$ as a function of temperature for $J_\text{K}=0.34$ (dashed line in (a)).  (c) Evolution of the probabilistic distribution $p(  V_{1})$ on the complex plane of $V_1=(V_1^x, V_1^y)$ at $T=0.001$ for three different values of $J_\text{K}$ in the magnetic regime, near the quantum critical point, and in the Kondo regime, respectively.}
\label{fig1}
\end{figure}

The above action is, however, generally unsolvable. To proceed, we propose a static approximation assuming $V_{i,n-m}=V_{i}\delta_{nm}$, $\chi_{n-m}=\chi\delta_{nm}$ such that $O_{nm}=O \delta_{nm}$. This ignores the temporal fluctuations of the auxiliary fields but takes full account of their spatial fluctuations and probabilistic distribution beyond the mean-field approximation \cite{Mukherjee2014PRB,Pradhan2015PRB,Karmakar2016PRA,Patel2017PRL,Jana2020JPCM,Karmakar2020JPCM}. To see how it works, we first integrate out all fermionic degrees of freedom and obtain an effective action only of the auxiliary fields,
\begin{equation}
S_{\text{eff}}=\beta\sum_{i=1}^{2}(\frac{J_{\text{K}}\left\vert V_{i} \right\vert ^{2}}{2}+\frac{J_{\text{H}}\left\vert \chi\right\vert ^{2}} {4}-\lambda_{i})-2\sum_{n}\ln\det (O-i\omega_n).
\end{equation}
The summation over Matsubara frequency can be evaluated using $\sum_{n}\ln\det(O-\operatorname*{i} \omega_{n})  =\sum_{l}\ln(  1+\operatorname{e}^{-\beta\xi_{l}})$, where $\xi_l$ denote the eigenvalues of $O$ and are always real because $O$ is Hermitian. This action can also be derived from an effective Hamiltonian (with the constraints $n^i_f=1$): 
\begin{align}
H_{\text{eff}}  \, &=\sum_{\mathbf{k}\sigma}\epsilon_{\mathbf{k}}c_{\mathbf{k}\sigma}^{\dag }c_{\mathbf{k}\sigma}+\frac{J_{\text{K}}}{2}\sum_{i,\sigma}\left( V_{i}c_{i\sigma}^{\dagger}f_{i\sigma}+\text{H.c.}\right)\nonumber\\
&\quad +\frac{J_{\text{H}}}{2}\sum_{\sigma}\left(  \chi f_{1\sigma}^{\dagger }f_{2\sigma}+\text{H.c.}\right),
\end{align}
where the auxiliary fields $V_i$ and $\chi$ are random variables satisfying $p_0(V_i)\sim \exp(-\beta J_{\text{K}}\vert V_{i} \vert^{2}/2)$ and $p_0(\chi)\sim \exp(-\beta J_{\text{H}}\vert \chi\vert^{2}/2)$, respectively. The model can then be studied using Monte Carlo simulations  \cite{Farkasovsky2010PRB,Kato2010PRL,Ishizuka2012PRL,Yang2019PRB,Maska2020PRB}.
Alternately, one may first eliminate the conduction electron part in $O$ and obtain $\sum_{n}\ln\det(O-\operatorname*{i} \omega_{n})=\sum_{n}\ln\det(A_n-\operatorname*{i} \omega_{n})+S_0$ with
\begin{equation}
A_n=\left[
\begin{array}
[c]{cc}
\lambda_{1}-\Delta_{11}(\operatorname*{i} \omega_{n}) & \frac{J_{\text{H}}\chi
}{2}-\Delta_{12}(\operatorname*{i} \omega_{n})\\
\frac{J_{\text{H}}\overline{\chi}}{2}-\Delta_{21}(\operatorname*{i} \omega_{n}) & \lambda_{2}%
-\Delta_{22}(\operatorname*{i} \omega_{n})%
\end{array}
\right].
\end{equation}
Here $\Delta_{ij}(\operatorname*{i} \omega_{n})=\frac{J_\text{K}^2\overline{V}_{i} V_{j}}{4N_{0}}\sum_{\mathbf{k}}\frac{\operatorname{e}^{-\operatorname*{i}\mathbf{k\cdot (\mathbf{R}_{i}-\mathbf{R}_{j})} }}{-\operatorname*{i}\omega_{n}+\epsilon_{\mathbf{k}}}$. $S_0$ is a constant from conduction electrons and can be safely dropped. The result is also real because $A_n^\dagger=A_{-n}$. 

The hybridization physics can then be studied with the probabilistic distribution $p( V_i, \chi)  =Z^{-1}\exp(  -S_{\text{eff}})$, where $Z$ is the partition function serving as the normalization factor. $V_i$ and $\chi$ are both complex numbers and, due to the high dimensionality of all variables, one may use the Monte Carlo and Metropolis algorithm for importance sampling. Compared to the perturbation expansion or mean-field theory, all spatial fluctuations of the hybridization fields are included. Below we will fix $J_\text{H}=0.1$ and the impurity distance $| \mathbf{R}|=| \mathbf{R}_2- \mathbf{R}_1|=1$ unless noted. For simplicity, we also set the half conduction bandwidth to unity and use $\epsilon_{\mathbf{k}}= -( \cos k_x +\cos k_y )/2$ with the particle-hole symmetry. The Lagrangian multipliers are then approximated by their saddle-point value $\lambda_i=0$ \cite{Saremi2011PRB}.

To better understand the fluctuation effect, we present in Fig.~\ref{fig1}(a) the mean-field phase diagram with three distinct regions. For small $J_\text{K}$, the mean-field hybridization is zero ($|V|=0$) and there is an artificial second order phase transition from $|\chi|=0$ to $|\chi|\neq 0$ at $T=J_\text{H}/4$, below which the two impurity spins form a singlet due to the Heisenberg  interaction. With increasing $J_{\text{K}}$, the mean-field theory predicts an artificial first-order transition to the Kondo phase ($|V|\neq0$) \cite{Jones1991PhysicaB,Jones1989PRB2}. By contrast, NRG and CFT analyses suggest a ``Varma-Jones" unstable fixed point at zero temperature \cite{Jones1987PRL,Jones1988PRL,Jones1989PRB1,Affleck1992PRL,Affleck1995PRB}  and argue that fluctuations will suppress the artificial first-order transition and  turn it into a crossover.

This is indeed the case in our method. The mean-field solution is equivalent to the saddle point approximation for the effective action $S_{\text{eff}}$. Figure~\ref{fig1}(b) compares the mean-field $|V_1|$ and that of maximal $p( V_1, V_2, \chi)$ at a chosen $J_{\text{K}}$ (the dashed line in Fig.~\ref{fig1}(a)). As expected, the two agree well with each other. However, if we first  integrate out $\chi$ and estimate $|V_1|$ from the maximum of the joint distribution $p( V_1, V_2)$, the transition temperature will be greatly suppressed. Since $p( V_1, V_2)$ includes the probabilistic distribution of $\chi$, the difference reflects the effect of fluctuating magnetic correlations between impurities in reducing Kondo screening by conduction electrons. We will show later that once the statistical fluctuations of $V_i$ are also included, the first-order transition does turn into a crossover in physical properties. Here just for illustration, we plot in Fig.~\ref{fig1}(c) the distribution $p(V_1)$ of the complex field $V_1=(V_1^x, V_1^y)$ after integrating out $V_2$ and $\chi$ from $p( V_1, V_2, \chi)$. With increasing $J_{\text{K}}$, the region of  large $p(V_1)$ is seen to first expand from a spot around the origin and gradually develop into a ring, showing a ``Mexican hat" potential for the hybridization in the Kondo phase.

To extract useful information on hybridization fluctuations, we define the mutual information of two random variables $X$ and $Y$ with the joint probability $p( x,y)$,
\begin{equation}
I(  X;Y)  =\iint p(  x,y)  \log \frac{p( x,y)  }{p(  x)  p(  y)  }dxdy, \label{MIDefinition}
\end{equation}
where $p(  x)  =\int p(  x,y)dy$ and $p(  y)=\int p(  x,y)dx$ are the marginal probabilities. Its calculation has historically been challenging because we typically only have samples rather than the underlying distribution \cite{Poolel2019PMLR}. A straightforward approach is to partition the samples into bins of finite size but the results are very sensitive to the bin sizes \cite{Kraskov2004PRE}. Here we use a neural estimator recently proposed for the mutual information \cite{Belghazi2018PMLR}:
\begin{equation}
I_\theta(X;Y) = \sup_{\theta \in \Theta} \mathbb{E}_{p\left( x,y\right)}[f_\theta] - \log \left(\mathbb{E}_{p\left( x\right) p\left( y\right)}[e^{f_\theta}] \right),
\end{equation}
where $f_\theta$ is a function parametrized by neural networks with the parameters $\theta \in \Theta$. $I_\theta$ gives a lower bound for the true mutual information, $I(X;Y) \geq I_\theta(X;Y) $, following the Donsker-Varadhan representation \cite{Donsker1975CPAM}. As long as the parameter space $\Theta$ is large enough, the inequality is tight and becomes a good approximation. The neural estimator has been successfully applied to both thermal and athermal systems \cite{Nir2020PNAS}. We implement it here with a three-layer neural network using Tensorflow and the Adam optimizer \cite{TensorFlow16,Kingma2015ICLR}.

\begin{figure}[ptb]
\begin{center}
\includegraphics[width=8.6cm]{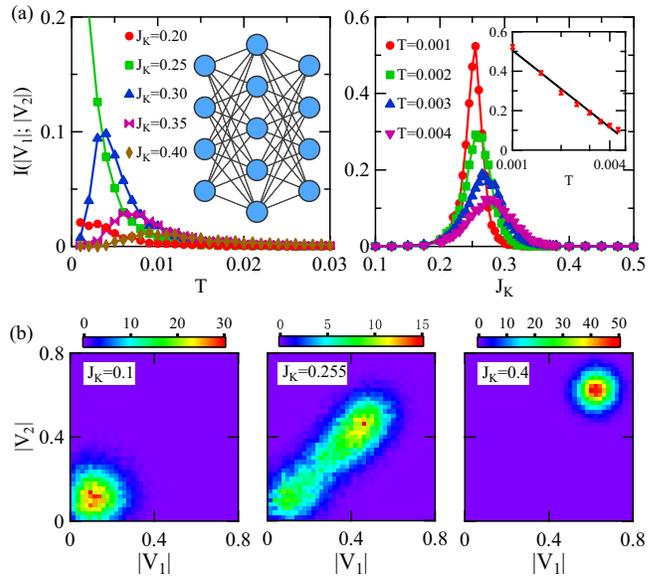}
\end{center}
\caption{(a) The amplitude mutual information $I(  |V_1|;|V_2|)$ as a function of the temperature and Kondo coupling. The left inset shows a schematic diagram of the three-layer neural network and the right inset shows logarithmic divergence of the maximal mutual information at low temperatures. The black solid line is a guide to the eye. (b) Comparison of the  distribution $p( |V_1|, |V_2| )$ at three different $J_\text{K}$ at $T=0.001$.}
\label{fig2}
\end{figure}

We first integrate out the phase $\theta_i$ of the complex $V_i=|V_i|\operatorname{e}^{\operatorname*{i}\theta_{i}}$ and study the mutual information of the amplitude $|V_i|$. The results are plotted in Fig.~\ref{fig2}(a) as a function of the temperature and Kondo interaction, respectively. For small $J_{\text{K}}$, the amplitude mutual information $I(  |V_1|;|V_2|)$ is small but grows with lowering temperature due to increasing intersite magnetic correlations; while for large $J_\text{K}$, a peak appears at finite temperature, indicating the weakening of intersite magnetic correlations due to Kondo screening at low temperatures. Remarkably, the amplitude mutual information varies nonmonotonically with $J_\text{K}$ and exhibits a maximum whose height increases rapidly and diverges logarithmically with lowering temperature, manifesting the quantum critical behavior above the ``Varma-Jones" fixed point. To get an intuitive picture, we compare in Fig.~\ref{fig2}(b) the distribution $p( |V_1|, |V_2|)$ at three different $J_\text{K}$. We see that the values of $|V_i|$ are scattered along the diagonal direction near the critical $J_\text{K}$. Hence the divergence comes from strong cooperative fluctuations of the hybridization fields on two impurities. Note that the maximal distribution of $|V_i|$ is not around the origin for small $J_\text{K}$ in polar coordinate, showing the presence of hybridization fluctuations even deep inside the magnetic phase.

\begin{figure}[ptb]
\begin{center}
\includegraphics[width=8.6cm]{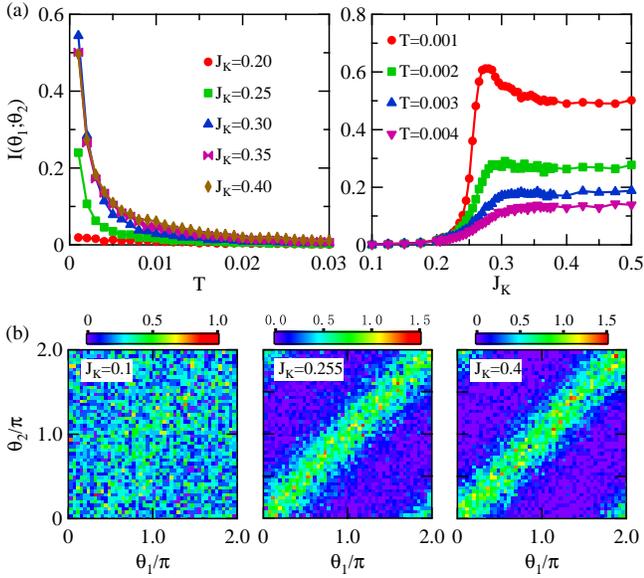}
\end{center}
\caption{(a) The phase mutual information $I(  \theta_1;\theta_2)$ as a function of the temperature and Kondo coupling. (b) Comparison of the distribution $p( \theta_1,\theta_2 )$ for different $J_\text{K}$ at $T=0.001$.}
\label{fig3}
\end{figure}

The mutual information of the phases is presented in Fig.~\ref{fig3}(a). Since the effective action of the two-impurity model is invariant under the transformation $V_{i}\rightarrow V_{i}\operatorname{e}^{\operatorname*{i}\phi_{i}}$, $\chi\rightarrow\chi\operatorname{e}^{-\operatorname*{i}(  \phi_{1}-\phi_{2})  }$, we fix the gauge such that $\chi$ is real and nonnegative and study the phase mutual information $I(  \theta_1;\theta_2)$ of the hybridization fields. We see that it is nearly zero at high temperatures where the intersite magnetic correlations are negligible. But unlike that of the amplitude, here it always grows with decreasing temperature and, in the Kondo regime, varies only slightly with $J_\text{K}$ as shown in Fig.~\ref{fig3}(a). Its large value reflects the establishment of cooperative phase fluctuations between two Kondo impurities. To see this more clearly, we plot in Fig.~\ref{fig3}(b) the probabilistic distribution $p(\theta_1,\theta_2)$ for three different values of $J_\text{K}$. For small $J_\text{K}$, the distribution is featureless; while for large $J_\text{K}$, it is peaked along the diagonal direction, indicating that the two are ``locked" with each other. The mutual information therefore provides a useful tool to measure the intersite phase coherence of the hybridization fields. Note that while the pattern of $p(\theta_1,\theta_2)$ may shift with the gauge of $\chi$, $I(  \theta_1;\theta_2)$ is gauge invariant. It is also related to the Shannon entropy of the phase difference through $I(  \theta_{1};\theta_{2})  =\text{const}-S(  \theta)$ where $S(  \theta)  =-\int p(  \theta)  \log p(\theta)d\theta $ with $\theta=\theta_1-\theta_2$.

\begin{figure}[ptb]
\begin{center}
\includegraphics[width=8.6cm]{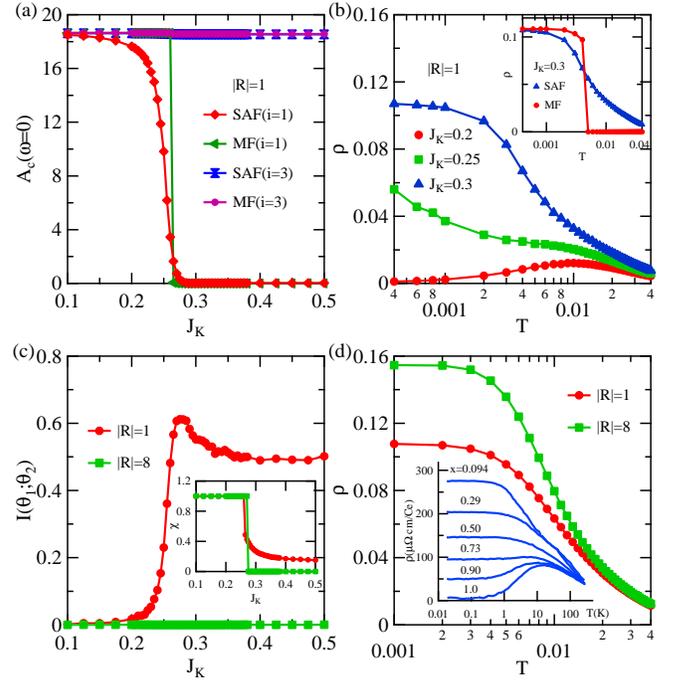}
\end{center}
\caption{(a) Comparison of the our results (SAF) and the mean-field results (MF) for the electron spectrum at ($i=1$) and away from ($i=3$) the impurity sites at $T=0.001$, showing strong suppression (pseudogap) due to hybridization fluctuations even in the magnetic regime.
(b) Temperature evolution of the resistivity $\rho$ at three different $J_{\text{K}}$ compared with the MF result (inset).
(c) Comparison of the phase mutual information for different impurity distances $|\mathbf{R}|=1$ and 8 at $T=0.001$. The inset shows the mean-field order parameter $\chi$.
(d) Resistivity as a function of temperature for $J_{\text{K}}=0.4$, showing a significant drop due to the intersite phase coherence for $|\mathbf{R}|=1$ compared to that of the single impurity limit ($|\mathbf{R}|=8$). The inset reproduces the experimental resistivity of Ce$_{1-x}$La$_x$Cu$_6$ for various $x$ \cite{Onuki1987JMMM}.}
\label{fig4}
\end{figure}

We may further study physical properties under this scheme taking partly account of the effect of magnetic and hybridization fluctuations. Figure~\ref{fig4}(a) plots the local density of states (LDOS), $A_{c}(  i,\omega=0)  \approx -\frac{\beta}{\pi}G_{c}(  i,\tau =\beta/2)$, where $G_{c}(  i,\tau)  =-\langle \mathcal{T}_{\tau}[ c_{i}(  \tau)  c_{i}^{\dag}(  0)  ] \rangle$ is the Green's function of conduction electrons. We find a strong suppression (pseudogap) at the impurity sites in the Kondo regime, in agreement with that observed in the scanning tunnelling experiments \cite{Spinelli2015NC}. Interestingly, the suppression already starts in the magnetic regime owing to the hybridization fluctuations. There appears no abrupt change across the critical point, in contrast to the mean-field expectation of a first-order phase transition. Figure~\ref{fig4}(b) plots the magnetic resistivity as a function of temperature calculated using $\rho\approx\frac{-\pi T^{2}}{g_{xx}(  \tau=\beta/2)  }$, where $g_{xx}(  \tau)  =-\langle \mathcal{T}_{\tau}j_{x}( \tau)  j_{x}(  0)  \rangle$ is the current-current correlation function and $j_{x}(  \tau)  =\frac{\operatorname*{i}}{2}\sum_{l\sigma}[ c_{l+x,\sigma}^{\dag}c_{l,\sigma}-\text{h.c.}]$ is the current operator  \cite{Wei2017SciRep,Huang2019Science}. We find logarithmic divergence near the critical point, suppression in the magnetic regime and saturation in the Kondo regime, indicating non-Fermi liquid, metallic, and Kondo-like behaviors, respectively. Again, the smooth evolution across the mean-field phase boundary implies a crossover rather than a first-order phase transition once fluctuations are included as in our method.

The importance of phase coherence is more evidently seen if we increase the distance between two impurities. Figure~\ref{fig4}(c) compares the phase mutual information for $|\mathbf{R}|=1$ and $8$. For large distance $|\mathbf{R}|=8$, the mutual information is reduced to zero even in the Kondo regime. A large finite phase mutual information only appears for small $|\mathbf{R}|$, indicating a fundamental difference of the multi-impurity Kondo physics from the single-impurity case. For comparison, the mean-field order parameter $\chi$ is also shown in the inset and found to be zero for $|\mathbf{R}|=8$ and finite for $|\mathbf{R}|=1$. The phase coherence is therefore closely associated with the intersite magnetic correlations \cite{Nakatsuji2002,Yang2008}. As shown in Fig.~\ref{fig4}(d), it causes a significant drop of the resistivity for $|\mathbf{R}|=1$, in qualitative agreement with the experimental observation in Ce$_{1-x}$La$_x$Cu$_6$ at large $x$ \cite{Onuki1987JMMM}.

We conclude that the static auxiliary field approximation can take good account of some fluctuation effects beyond the mean-field approximation. The employment of the mutual information can effectively reduce the dimensionality of the data and allow one to extract key information hidden in the complicated probabilistic distribution functions. It is thus a useful tool to probe the quantum phase transition and phase coherence in Kondo systems. Our method can be extended easily to other Kondo models to investigate spatial correlations of the hybridization physics. In the dense Kondo lattice, we expect that it will also suppress the artificial phase transitions predicted in an effective Ginzburg-Landau-Wilson theory \cite{Ohara2007PRB,Ohara2013JPSJ,Ohara2014JPSJ} and reveal the importance of cooperative hybridization fluctuations and intersite phase coherence, which is important for understanding the heavy Fermi liquid \cite{Raczkowski2019PRL}. The method may also be used to reveal the snapshot or spatial modulation of hybridization configurations \cite{Zhu2008PRL,Pepin2011PRL,Thomson2016PRB} and provide novel insight on the Kondo physics in multi-impurity or lattice systems.

This work was supported by the National Key Research and Development Program of MOST of China (Grant No. 2017YFA0303103), the National Natural Science Foundation of China (NSFC Grant No. 11974397,  No. 11774401), the Strategic Priority Research Program of the Chinese Academy of Sciences (Grant No. XDB33010100), and the Youth Innovation Promotion Association of CAS.

\end{document}